\documentstyle[seceq,epsf,wrapft]{ptptex}

\newcommand{\simleq}{\; \raisebox{-0.4ex}{\tiny$\stackrel
{{\textstyle<}}{\sim}$}\;}

\title{Spin Physics with the STAR detector}

\author{
L.C. {\sc Bland}\footnote{E-mail address: bland@iucf.indiana.edu}, for
the STAR collaboration
}

\inst{
Department of Physics, Indiana University, Bloomington, IN   USA}

\abst{
The STAR detector will bring unique capabilities to the study of
$\vec{p}+\vec{p}$ collisions up to total center of mass energies of
500 GeV at RHIC.  The large acceptance of the time projection chamber
and the electromagnetic calorimeter enable STAR to observe jets, and
to detect photons and electrons in important regions of phase space.
The premier experiments of the STAR spin program are discussed in this
contribution.  Photon + jet coincidence measurements will provide the
world's best determination of the contribution gluons make to the
proton's spin.  Detection of the daughter electrons and positrons from
$W^\pm$ decay will enable the determination of the polarization of the
$q\overline{q}$ sea within the proton.
}

\begin{document}

\maketitle

\section{Introduction}

\indent

With the addition of Siberian Snakes and spin rotators, the
Relativistic Heavy Ion Collider (RHIC) at Brookhaven National
Laboratory will have the unique capability of accelerating intense
polarized proton beams up to energies of 250 GeV per beam.  A series
of experiments studying $\vec{p}+\vec{p}$ collisions at energies
up to $\sqrt{s}$ = 500 GeV will then be possible at the PHENIX and
STAR detectors.  The luminosities expected for the RHIC spin program
(up to $2 \times 10^{32} {\rm cm}^{-2}{\rm s}^{-1}$ at $\sqrt{s}$ =
500 GeV) will enable the study of {\it polarization observables} for
large transverse momentum ($p_T$) processes, where perturbative QCD
methods have been successfully applied to explain most {\it
unpolarized} observables in $p+\overline{p}$ collisions.  One of the
polarization observables that will be measured at RHIC is the
parity-allowed longitudinal spin correlation
$$A_{LL}={1 \over P^2}{N_{++} - N_{+-} \over N_{++}+N_{+-}},\eqno(1)$$
where $P$ represents the beam polarizations assumed to be the same for
both longitudinally polarized beams in Eqn. 1.  The RHIC-spin project
will theoretically provide $P=0.70$ for each beam.  The spin
correlation coefficient, $A_{LL}$, requires measuring the difference in
yield for some process for equal ($N_{++}$) and
opposite ($N_{+-}$) proton beam helicities.  Parity-violating
single-spin longitudinal asymmetries ($A_L$) have a similar form to
Eqn.~1, but require that only one beam be polarized.

This contribution discusses the capabilities the STAR detector brings
to the RHIC-spin program.  When the construction of the barrel and
endcap electromagnetic 
calorimeters are completed, STAR will have unique capabilities for the
study of large-$p_T$ processes in $\vec{p}+\vec{p}$ collisions.  In
addition to critical tests of QCD, the STAR detector will study
processes that will enable the untangling of the spin structure of the
proton.  In particular, the study of the
$\vec{p}+\vec{p}\rightarrow\gamma+{\rm jet}+X$ reaction at $\sqrt{s}$=200 and
500 GeV will provide the world's best determination of the fraction of
the proton's spin carried by gluons.  Measurement of polarization
observables for inclusive jet and di-jet production will provide
important cross checks on the gluon polarization.
In addition, the study of parity
violating spin asymmetries in $W^\pm$ production will help unravel
the spin-flavor content of the $q\overline{q}$ sea within the proton.
These spin structure function studies are the next essential step in
understanding the internal structure of the proton.

The present understanding of the proton's spin structure comes from
years of polarized deep-inelastic scattering (pDIS) experiments at CERN,
SLAC and more recently the HERMES experiment at HERA \cite{Petratos}.
A global analysis of the inclusive pDIS data suggests that the quarks
account for only a small fraction of the proton's spin, assuming that
the gluon polarization is zero.  This deficiency can be rectified if
the gluon polarization is large, and there are canceling
contributions to the proton spin arising from orbital angular
momentum.  The crucial next step in understanding the proton's spin
structure is to measure the gluon polarization.  The goal is the
determination of the {\it integral} $\Delta G$, defined as
$$\Delta G(Q^2) = \int_0^1 \Delta G(x,Q^2) dx = 
\int_0^1 [ G^+(x,Q^2)-G^-(x,Q^2)] dx, \eqno(2)$$
related to the total contribution gluons make to the proton's spin, at
the scale $Q^2$.  (Hereafter, the $Q^2$ dependence of
these quantities will be suppressed for clarity.)  The fraction of the
proton's longitudinal momentum carried by the gluon is $x$.
In Eqn.~2, the gluon helicity asymmetry distribution ($\Delta
G(x)$) is given by the difference in probability of finding a gluon
with its polarization parallel ($G^+$) versus 
antiparallel ($G^-$) to the proton's longitudinal polarization.  The
unpolarized gluon probability distribution is defined as
$G(x)=G^+(x)+G^-(x)$.  The gluon polarization at a given $x_{gluon}$ is
$\Delta G(x)/G(x)$.

Several ideas exist for probing the gluon polarization.  Although
gluons are not electrically charged, they can be directly probed in polarized
lepton scattering experiments using the so-called photon-gluon fusion (PGF)
process.  The COMPASS experiment \cite{COMPASS} 
will attempt to study this process by
detecting the leading hadrons from the $q$ and $\overline{q}$ jets produced
in PGF.  The range of $x_{gluon}$ covered by COMPASS will prevent a
determination of the integral in Eqn.~2.
COMPASS will have sensitivity to the
gluon polarization at relatively large
$x_{gluon}$, and hence will be insensitive to the $x$ range where the
gluons become increasingly abundant.  Even though the gluon
polarization is small at small $x$, there are critical contributions
to the integral in Eqn.~2, as $x\rightarrow 0$.  The best
determination of the integral $\Delta G$ requires sensitivity to the gluon
polarization over a broad range of $x_{gluon}$.

An alternate method for determining $\Delta G(x)$ is to study
pDIS over a very broad range of $x_{quark}$ and $Q^2$.  The gluon helicity
asymmetry distribution could then be extracted by an analysis of the
scaling violations in pDIS.  Existing measurements of pDIS do not span
a broad enough range of $Q^2$ to provide an accurate
determination of $\Delta G(x)$.  The necessary
pDIS measurements could be
made at HERA in the study of $\vec{e}+\vec{p}$ collisions, if
the 820 GeV proton beam were polarized.  A decision to
take this challenging step will not be made until 2002.

In the absence of a `polarized HERA collider', the best determination
of the integral $\Delta G$ will come from the study of photon production in
$\vec{p}+\vec{p}$ collisions at RHIC, as described below.

\section{The STAR detector}

STAR is one of the two large detectors at RHIC designed for
relativistic heavy-ion collisions to isolate and study the
quark-gluon plasma.  As is also true for PHENIX, the capabilities of
the STAR detector will allow its use for the study of many of the
interesting processes in $\vec{p}+\vec{p}$ collisions.  The large
acceptance of the STAR detector makes it suitable for reconstructing
the hadronic jets that are prolifically produced in high-$p_T$
processes.  The STAR electromagnetic calorimeter will be used to
detect high-$p_T$ photons and the $e^\pm$ daughters of the $W^\pm$
bosons produced in $\vec{p}+\vec{p}$ collisions.  In addition,
the $e^+e^-$ pairs produced in either the Drell-Yan or real $Z^0$
production can be detected with the large acceptance of the STAR EMC.

The heart of the STAR detector is the world's largest time projection
chamber.  The TPC provides tracking through a 0.5 T solenoidal
magnetic field of the thousands of charged particles that are
produced in a single Au-Au collision.  The TPC will play an important
role in the STAR-spin program, tracking the charged hadrons produced
by jets over the pseudorapidity interval $-1.8\simleq\eta\simleq 1.8$.
Photons and electrons will be detected by the STAR electromagnetic
calorimeter (EMC).  The barrel EMC construction is well underway, and
will ultimately result in 120 modules spanning $2\pi$ in azimuth and the
pseudorapidity interval, $-1 \le \eta \le +1$.  At present, a patch of
the BEMC consisting of four modules is installed in the STAR
detector.  The patch covers $0 \le \eta \le +1$, with an azimuthal
extent of 0.42 radians.  The BEMC patch will be in operation for the
first Au-Au collisions in the year 2000.

\begin{figure}
\epsfxsize=10 cm
\centerline{\epsfbox{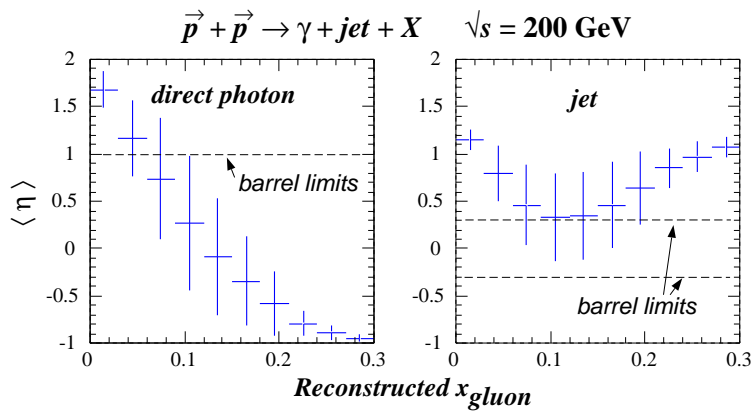}}
\caption{{\it  Moments of the pseudorapidity distribution of simulated
photon-jet coincidence events contributing to each bin in
reconstructed $x_{gluon}$ values.  Each point represents the mean
contributing $\eta$ value for photons or jets, with the `error' bars
reflecting the rms deviation from the mean.}}
\label{gam_jet_kin}
\end{figure}

Recently, funding for
the STAR endcap EMC was approved.  When its construction is
complete, the EEMC will span the pseudorapidity interval, $1.07 \le
\eta \le 2$, covering the full azimuth, and will provide access to
{\it asymmetric
partonic collisions} for `direct photon' and $W^\pm$ production
processes.  The EEMC provides critical phase space coverage for both
$\gamma$+jet and $W^\pm$ production studies.  By detecting large-$p_T$
processes at forward angles, asymmetric initial states, where one
parton has a larger $x$ than the other, are emphasized.  For photon
production, such collisions effectively select large-$x$ quarks as an {\it
analyzer} of the polarization of small-$x$ gluons.  In addition to
kinematic selection of quarks with large polarization (the pDIS
asymmetry $A_1^p$ increases with increasing $x_{quark}$), the EEMC
will detect photons produced at partonic CM angles where the
partonic-level spin correlation parameter ($\hat{a}_{LL}$) approaches
its maximum value, unlike the situation encountered for midrapidity
photon production.

The large acceptance of STAR is critical for the measurement of the
away-side jet in coincidence with the produced photon 
(Fig.~\ref{gam_jet_kin}).  Coincident
$\gamma$+jet detection allows for the reconstruction of the
initial-state partonic kinematics.  With the momentum fractions
$x_{quark}$ and $x_{gluon}$ determined by the experiment, a more
direct connection between the measured polarization observables and
the gluon helicity asymmetry distribution can be made.  This is
advantageous to in trying understand how the measurement errors will influence
the determination of $\Delta G(x)$.

\section{Gluon polarization measurements at STAR}

\begin{wrapfigure}{r}{6.6cm}
\epsfxsize=6.5 cm
\centerline{\epsfbox{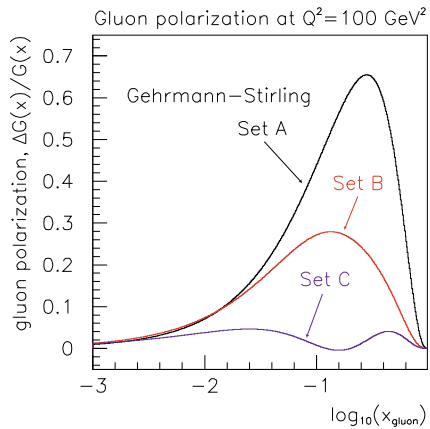}}
\caption{{\it  Gluon polarizations computed from models of $\Delta
G(x)$ consistent with polarized deep inelastic scaling violations
\cite{GS}.  The structure functions are evolved to the scale that
will be probed at RHIC.}}
\label{gluon_pol}
\end{wrapfigure}

The existing data for scaling violations in pDIS
provide only very loose constraints on $\Delta G(x)$.
Several analyses of these constraints have been made \cite{BFR96,GS}
and have generally concluded that the {\it integral} $\Delta G$ 
is positive.  The variation of the gluon helicity asymmetry
distribution with the gluon momentum fraction ($x_{gluon}$) 
has significant differences
in these different analyses.  There is generally always a
positive peak of $x\Delta G(x)$, but the $x_{gluon}$ value of the peak
is not well constrained.  Consequently, the gluon {\it polarization},
($\Delta G(x) / G(x)$) can be either large or
small, depending on where the peak in $x\Delta G(x)$ occurs.  Many
parameterizations of $\Delta G(x)$ that are consistent with 
existing measurements result in negatively polarized gluons at some
$x_{gluon}$ values.  In leading
order perturbative QCD (pQCD), the spin correlation parameter ($A_{LL}$) that
will be measured in $\vec{p}+\vec{p}\rightarrow \gamma + X, \gamma$ +
jet + $X$ reactions at RHIC is proportional to the gluon polarization.  

To illustrate the sensitivity of $\gamma$ + jet coincidence
measurements planned for STAR, simulations using the three $\Delta
G(x)$ models in Ref. \cite{GS} (hereafter referred
to as GS sets A,B and C) have been performed.  In all cases, the
input $\Delta G(x)$ must be evolved \cite{Kumano} from the scale where
the analysis was performed, $Q_0^2$=4 GeV$^2$, to the scales
that will be probed at RHIC, taken to be $Q^2$=$p_{T,\gamma}^2/2$,
where $p_{T,\gamma}$ is the transverse momentum of the photon.
The variation of the resulting gluon polarization with $x_{gluon}$ for the
three $\Delta G(x)$ models in Ref. \cite{GS}, evolved to $Q^2$=100
GeV$^2$, is shown in Fig.~\ref{gluon_pol}.

\begin{figure}
\epsfxsize=10 cm
\centerline{\epsfbox{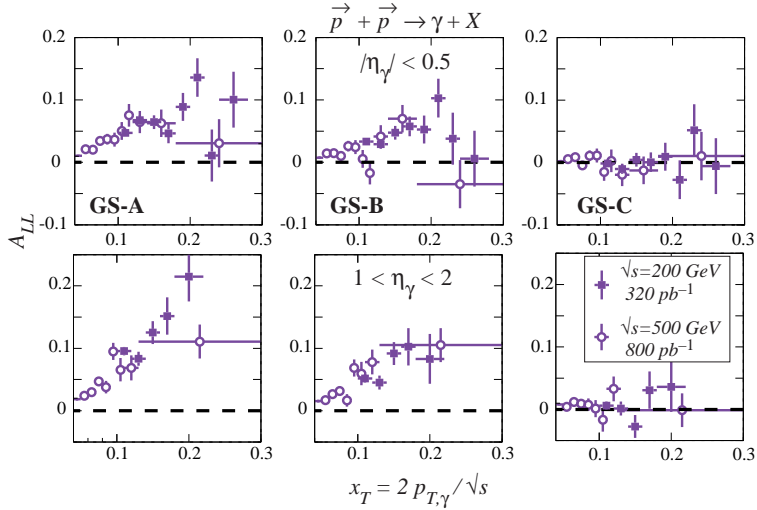}}
\caption{{\it  Simulated values for $A_{LL}$ for inclusive photon
production at RHIC energies.  The top row shows the spin correlations
for midrapidity photons that can be detected at both PHENIX and STAR.
The bottom row shows expected values at the forward angles probed by
the STAR EEMC.  Not evident in this figure, is that for a given $x_T$,
photons detected in the EEMC correspond to smaller-$x$ gluons than
those detected at midrapidity.}}
\label{ALL_incl_gam}
\end{figure}

The simulated spin correlation coefficient \cite{EPIC} expected for
inclusive photon production in $\vec{p}+\vec{p}$ collisions at RHIC, using the
GS-A,B and C models of $\Delta G(x)$, is shown in
Fig.~\ref{ALL_incl_gam} as a function of $x_T=2 p_{T,\gamma}/\sqrt{s}$.
The variable $x_T$ is often interpreted as the initial-state parton
momentum fraction.  This applies only for midrapidity photons, and is
only true when averaging over an event ensemble.  The
simulations include the subprocesses $fg \rightarrow f\gamma$ (where
$f$ refers to either a quark or an antiquark), $q\overline{q} \rightarrow
g\gamma$ and $q\overline{q} \rightarrow \gamma \gamma$, hereafter
referred to as `direct photon' processes.  The first of
these subprocesses provides $\sim$90\% of the direct photon yield.  The
magnitude of the calculated $A_{LL}$ in the pseudorapidity range covered by
the STAR EEMC is larger than at midrapidity, as expected since the endcap
selects events having more asymmetric $qg$ collisions and having partonic
scattering angles where $\hat{a}_{LL}$ (the pQCD result for the partonic
spin correlation coefficient) is large.  The predicted decrease in $A_{LL}$ as
one goes from GS set A, to set B and then set C for $\Delta G(x)$,
simply reflects the smaller gluon {\it polarization vs} $x_{gluon}$ in
those three models (Fig.~\ref{gluon_pol}).

\begin{figure}
\epsfxsize=10 cm
\centerline{\epsfbox{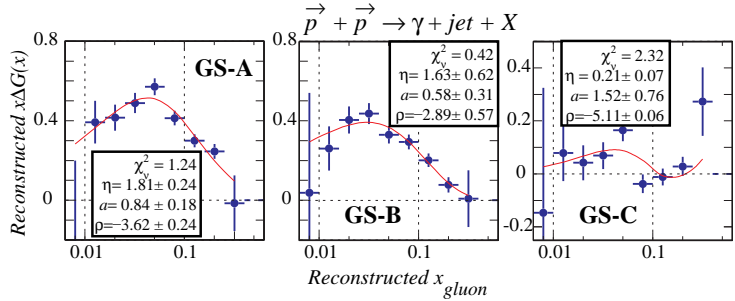}}
\caption{{\it  Fits to the reconstructed $\Delta G(x)$ directly
reconstructed from the simulated $A_{LL}$ for $\gamma$+jet
coincidences, after correcting for $q\overline{q}$ annihilation.  A
standard parameterization is used to fit the data, 
with the parameters
specifying the large $x_{gluon}$ behavior fixed.  Full data sets for
$\sqrt{s}$=200 (500) GeV are assumed, corresponding to integrated
luminosity of 320 (800) pb$^{-1}$}}
\label{direct_extr}
\end{figure}

As discussed elsewhere \cite{EPIC,CDR}, STAR will be able to
detect a significant fraction of the away-side jets in coincidence
with the produced photon.  This capability results from the large 
phase space coverage of
the existing time-projection chamber and the planned electromagnetic
calorimetry.  Detection of $\gamma$+jet coincidences enables
the reconstruction of the initial-state partonic
kinematics \cite{EPIC}.  With this capability, a direct extraction of $\Delta
G(x)$ can be made from the measured $A_{LL}$, assuming contributions
from only quark-gluon Compton scattering and collinear initial-state
parton collisions.  Excluding experimental backgrounds, there remain small
contributions to the $\gamma$+jet yield from partonic subprocesses
other than quark-gluon Compton scattering.  The $q\overline{q}$
annihilation contribution can be corrected for based on simulations.
Fig.~\ref{direct_extr} shows the directly extracted $\Delta G(x)$ from the
simulated $A_{LL}$ values, after applying an additive correction for
$q\overline{q}$ annihilation.  A 320 pb$^{-1}$ sample at
$\sqrt{s}=200$ GeV and a 800 pb$^{-1}$ sample at $\sqrt{s}=500$ GeV
have been combined in the figure.  These data samples can be achieved
in two ten week runs, based on the projected luminosity for
$\vec{p}+\vec{p}$ collisions.  The latter is especially crucial to
extend the coverage to small $x_{gluon}$ values.  

To ascertain the
sensitivity STAR will have to the fraction of the proton's spin
carried by gluons (or, the {\it integral} $\Delta G$), the results in
Fig.~\ref{direct_extr} were fitted to a standard structure function
parameterization \cite{GS}, 
$$x\Delta G(x) = \eta A x^a (1-x)^b [1 + \rho x^{1/2} + \gamma x],$$
$$ {\rm with~~} A^{-1}=\left( 1+{\gamma a 
\over a+b+1}\right){\Gamma(a) \Gamma(b+1) 
\over \Gamma(a+b+1)}
+ \rho {\Gamma(a+{1 \over 2})\Gamma(b+1) \over \Gamma(a+b+1)}.  \eqno(3)$$
The resulting fitted value for $\eta$
represents the integral $\Delta G$.
In Eqn.~3, the $b$ and $\gamma$ parameters are held fixed at values
obtained by evolving \cite{Kumano} the $\Delta G(x)$ input to the
simulation to the $Q^2$ values relevant at RHIC.  
The $b$ parameter specifies the large-$x$ behavior of $\Delta G(x)$.
The fixed parameters used in the fits are consistent with positivity
constraints ($|\Delta G(x)| < G(x)$).  

Beyond illustrating the sensitivity of STAR to the fraction of the
proton's spin carried by gluons, the analysis presented in
Fig.~\ref{direct_extr} illustrates several other things.
\begin{itemize}
\item {\it an accurate determination of $\Delta G$ will require both
$\sqrt{s}$=200 and 500 GeV data samples to get to sufficiently small
$x_g$.}  Due to
strong correlations between $\eta$ and $a$ (specifying the small $x$
behavior of $\Delta G(x)$), $\delta \eta$ grows
rapidly as the low-$x$ points are successively eliminated.  It is
critical to observe the falloff of $x\Delta G(x)$ with decreasing $x$
to ensure an accurate determination of $\eta$.

\item {\it the large-$x$ behavior of $\Delta G(x)$ must be constrained to
determine the integral $\Delta G$ at RHIC}.  The fixed parameters in the above
analysis presuppose that the large $x$ behavior of $\Delta G(x)$ is
known.  The COMPASS experiment at CERN \cite{COMPASS} should 
provide the necessary measurements.  Detection of $\gamma$+jet
coincidences, with the photons observed at negative pseudorapidity 
(Fig.~\ref{gam_jet_kin}), will provide critical overlap between 
the STAR and COMPASS results to check
the consistency of $\Delta G(x)$ between the two experiments.

\item {\it Additional corrections must be made to $\Delta G_{recon}(x)$
to ensure an accurate determination of the integral $\Delta G$.}
The present analysis neglects several other corrections, including
evolving all of the $\Delta G_{recon}(x)$ points to a common $Q^2$ and
correcting for the kinematic reconstruction errors, because they
require knowledge of $\Delta G(x)$ and hence will require an
iterative approach to deduce the result.  Even when making these
corrections, the fitted $\eta$, although closer to the input value, is
too small.  The largest remaining error comes
from neglecting the transverse momentum ($k_T$) of the partons in the initial
state.  Repeating the analysis with simulations that 
don't include initial-state parton showers (and hence do not include
$k_T$ smearing) results in a fitted $\eta$ in agreement with the input
$\Delta G$. 
\end{itemize}

The end result is, that after accounting for the most significant
sources of systematic error \cite{CDR}, we expect that the fraction
of the proton's spin carried by gluons can be determined to an
accuracy of approximately 0.5, primarily based on the STAR measurements of
$\vec{p}+\vec{p}\rightarrow \gamma+{\rm jet}+X$.  Data samples at both
$\sqrt{s}$=200 and 500 GeV are crucial so that the accuracy is not
limited by extrapolation errors.  The analysis of $\Delta
G_{recon}(x)$ presented here is intended to illustrate the sensitivity
of the STAR measurements to the integral $\Delta G$.  Clearly, the best
determination of $\Delta G$ will result from a global analysis of all
relevant data.  

To achieve this accuracy in the integral $\Delta G$, the relatively
small $\gamma$+jet signal must be extracted from a sizeable background
of $\pi^0(\eta)$+jet coincidences, where a high-$p_T$ meson can mimic
the photon signal by decaying into two closely-spaced photons.
Both isolation cuts and observation of the characteristic response of a
shower-maximum detector to the showers from the closely spaced photons
will be used \cite{EPIC,CDR} to 
discriminate signal from background at STAR.  In
addition to these experimental backgrounds, other processes will
contaminate the quark-gluon Compton scattering signal.

\begin{figure}
\epsfxsize=7.5 cm
\centerline{\epsfbox{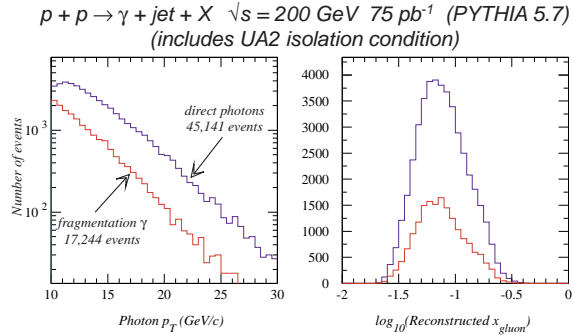}}
\caption{{\it  Comparison between the `direct' and `fragmentation'
$\gamma$ + jet yield that is expected to be observed with the STAR
detector.  Direct photons are those produced in hard scattering
events, predominantly quark-gluon Compton scattering.  Fragmentation
photons are produced in the fragmentation of recoiling final state
partons.  It is possible that a more restrictive isolation condition
will reduce the contribution from fragmentation photons.}}
\label{frag_dir}
\end{figure}

As predicted by several studies \cite{Vogel}, photons produced in the
fragmentation of final state recoiling partons present an important
contribution to the total yield, in addition
to `direct photon' processes.  The impact of these so-called
`fragmentation photons' on the determination of $\Delta G(x)$ 
using $\gamma$+jet
coincidences at STAR has been ascertained using PYTHIA \cite{SJ94}.  Reliable
results are expected from the simulated `fragmentation photon' yield, since the
$p+\overline{p}\rightarrow \gamma$ + 2 jet + $X$ yield measured by CDF
\cite{gam2jet} is well represented by PYTHIA. 

The `fragmentation photon' yield is obtained by considering all
$2\rightarrow2$ subprocesses, responsible for the bulk of the
non-diffractive inelastic cross section, and searching for those
events that have an energetic photon that is not produced by the decay
of any parent hadron.  The detector resolutions and acceptances of
STAR are imposed on all observable particles in the event.
The away-side jet is reconstructed using the UA1 jet finder, suitably
modified for STAR \cite{SN196},
and the UA2 \cite{UA2} isolation condition is imposed on the
high-$p_T$ photon candidate.
Naively, the latter
is expected to effectively eliminate `fragmentation photons', because of the
additional hadrons expected within a cone around the photon.
Unfortunately, the hardest fragmentation 
photons result from bremsstrahlung, and
are widely displaced from the core of the jet produced by the
radiating parton.  The end 
result is that a significant fraction of the overall $\gamma$ + jet
yield will arise from `fragmentation photons'.  The comparison between
this yield and the `direct photon yield' (defined in the previous
section) is 

\begin{wrapfigure}{r}{6.6cm}
\epsfxsize=6.5 cm
\centerline{\epsfbox{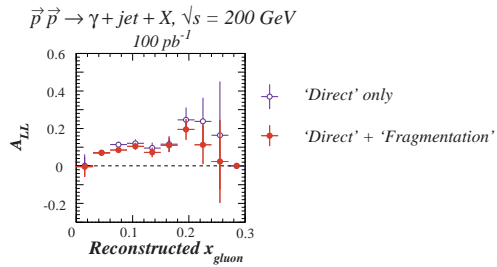}}
\caption{{\it  Calculated $A_{LL}$ for
$\vec{p}+\vec{p}$ collisions at $\sqrt{s}=200$ GeV for `direct
photons' only versus `direct + fragmentation' photon production.  
The sizeable
yield of `fragmentation photons' produces a small dilution of $A_{LL}$.}}
\label{spin_correlation}
\end{wrapfigure}
\noindent
shown in Fig.~\ref{frag_dir}.  It is possible that an improved
isolation condition, beyond the one employed by UA2, can reduce the
`fragmentation photon' yield beyond that shown in the figure.

What impact does this background process have on the planned
measurement of $\Delta G(x)$?  This question was addressed by
calculating $A_{LL}$ for $\vec{p}+\vec{p}$ collisions including
both `direct' and `fragmentation' photon processes.  
The result for
$A_{LL}$ is shown in Fig.~\ref{spin_correlation}, comparing `direct photon'
production alone to a calculation combining direct and fragmentation
photons.  The input $\Delta G(x)$ corresponds to GS set A \cite{GS}.
A small dilution of $A_{LL}$ is observed from
fragmentation photons.  Either improved isolation cuts or a more
sophisticated analysis of the event topology is expected to reduce
the dilution of the direct photon $A_{LL}$.  The relative importance
of the dilution will increase as the gluon polarization decreases.

\section{Partonic kinematics reconstruction in $W^\pm$ production}
\indent

The Standard Model predicts that $W^\pm$ bosons are predominantly
produced in $p+p$ collisions by the partonic processes
$u+\overline{d}\rightarrow W^+$ and $d+\overline{u}\rightarrow W^-$.  
Given the $V-A$ theory of
the weak interaction, sizeable {\it parity violating} longitudinal
spin asymmetries, $A_L$, are expected in $\vec{p}+p$ collisions \cite{BS93}.
These asymmetries can be related to the polarized and unpolarized
parton distribution functions, and in certain kinematic domains are
directly proportional to either quark or antiquark polarizations
\cite{CDR} ({\it
ie.}, the ratio of the structure functions $\Delta f(x)/f(x)$, where
$f$ represents either $q$ or $\overline{q}$). 
Since the partonic constituents of the proton are assumed to have
negligibly small transverse momenta, the produced $W^\pm$ bosons should
be collinear with the colliding protons.  Higher-order gluon radiation
\cite{AEM85} results in non-zero, but small, values for the $W^\pm$ transverse
momentum, defined as $q_{T}$.

To achieve the goal of determining the unpolarized and polarized parton
distribution functions of the nucleon, the variation of these
probabilities with Bjorken $x$, interpreted as the
fraction of the nucleon's longitudinal momentum carried by the parton,
and the scale, $Q^2$ is required.  In some cases, this $x$
dependence is deduced from a theoretical interpretation of experimental
observables.  The impact of the measurement errors on the deduced
structure functions can be best ascertained if these critical
kinematic variables can be {\it directly deduced} from the experiment.
For the case of determining the gluon polarization within the proton,
$\gamma$+jet coincidences provides the necessary kinematic
determination.  Below, a method is proposed to 
extract the $x$ dependence of
the polarization of sea antiquarks in the proton.
The method relies on measuring only the four momentum of the daughter
charged lepton produced in the decay of $W^\pm$, and simplifying
assumptions about the kinematics.

\begin{figure}
\epsfxsize=9.5 cm
\centerline{\epsfbox{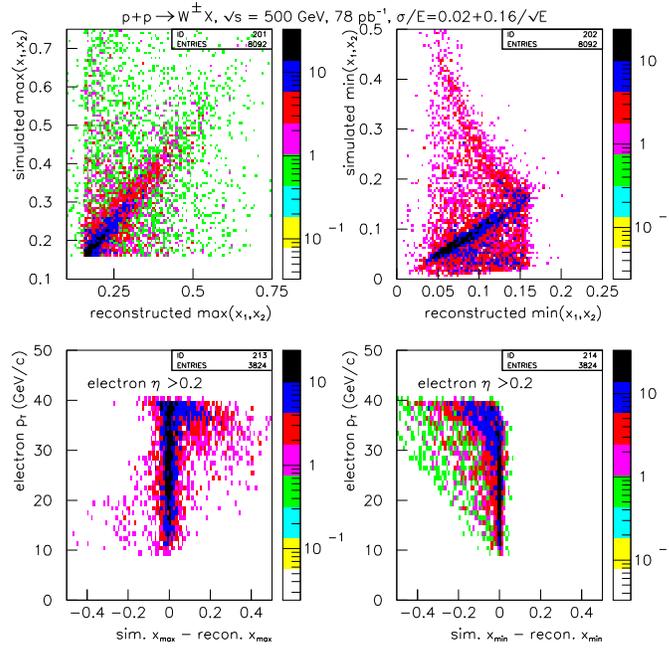}}
\caption{{\it  (Top) Simulated versus reconstructed Bjorken x values
for the $q$ and $\overline{q}$ that form the $W^\pm$ in $p + p$
collisions.  The reconstruction ignores the transverse momentum of the
$W$ ($q_T$).  (Bottom)  The difference between simulated and reconstructed $x$
value versus the transverse momentum of the daughter $e^\pm$ produced by
$W^\pm$ decay.  The $e^\pm$ are detected at $\eta > 0.2$ to minimize the
number of events were the $W^\pm$ has small longitudinal momentum ($p_{L,w}$).
The initial-state kinematics reconstruction fails at
midrapidity because $q_{T}$ is not small relative to $p_{L,W}$.}}
\label{sim_v_recon}
\end{figure}

If the $W^{\pm}$ transverse momentum is zero, then by simply measuring
the angle and energy of the daughter charged lepton, the longitudinal
momentum of the parent $W^\pm$ can be determined.  This, combined with
the assumption that the total energy in the partonic CM 
equals the nominal mass of the $W$, allows
the Bjorken $x$
values for the interacting $q$ and $\overline{q}$ to be directly
deduced.  In a real experiment, $q_{T}$ can be ignored if
it is small compared to the $W^\pm$ longitudinal momentum, $p_{L,W}$.
Conveniently, the region of phase space where $A_L$ can be most directly
related to the quark and antiquark polarization corresponds to
asymmetric $q + \overline{q}$ collisions, resulting in large
$p_{L,W}$.  When $p_{L,W}$ is large, $q_{T}$ can be assumed to be
zero.  The accuracy of this initial-state partonic kinematics
reconstruction is shown in Fig.~\ref{sim_v_recon}.  In that figure, 
an initial-state kinematics reconstruction is applied to $p+p
\rightarrow W^\pm$ events generated by PYTHIA \cite{SJ94}
at $\sqrt{s}$=500 GeV.  The parton shower
model is used to simulate higher-order QCD effects responsible for
non-zero $q_{T}$ values for the $W^\pm$.  Event selection requires that an
$e^+$ or $e^-$ be within the acceptance of the STAR barrel and endcap EMC and
have $p_{T,e} \ge 10$ GeV/c.  The resolution of the STAR EMC is
taken to be $\delta E_e/E_e = 0.02 + 0.16/\sqrt{E_e}$.
The finite accuracy in the
partonic kinematics reconstruction dominantly results from the simplifying
assumptions used in the analysis but also has some contributions from
the EMC resolution, imposing a performance requirement on the detector.

The critical assumption that $q_{T}$ is small works best when the
daughter $e^\pm$ from $W^\pm$ decay is detected away from midrapidity
($|\eta|\approx 0$).
The reason for this is that most of the $e^\pm$ at large $\eta$ are
produced from $W^\pm$ created in asymmetric $q + \overline{q}$
collisions \cite{CDR}.  These asymmetric collisions provide a sizeable
longitudinal momentum to the $W$,
$p_{L,W}=\beta_{pCM}\gamma_{pCM}M_{W}c$, where the partonic center of
momentum (pCM) is moving with velocity equal to $\beta_{pCM}c$ and
$\beta_{pCM} = (x_1-x_2) / (x_1+x_2)$ in the collider reference
frame.  The momentum fractions of the $q$ and $\overline{q}$ are
denoted as $x_1$ and $x_2$.  When the daughter $e^\pm$ are detected near
$|\eta|\approx 0$, the reconstruction procedure fails because
$p_{L,W}$ is generally small, and $q_{T}$ can no longer be ignored.
The influence of $q_{T}$ on the kinematics reconstruction can be
minimized by imposing restrictions on both $\eta_e$ and $p_{T,e}$.  The
correlation of $\delta x_{max(min)}$ with $p_{T,e}$, for 
events with $\eta_e>0.2$, is shown in
Fig.~\ref{sim_v_recon}.  The kinematics reconstruction is observed to
fail at the largest values of $p_{T,e}$.  
The extreme values of the $e^\pm$ transverse momentum are
known to be most sensitive to $q_{T}$ \cite{Abbott}.

Overall, this kinematics reconstruction procedure accurately
determines the momentum fractions of the quark and antiquark that form
the $W^\pm$ for most of the events.  However,
even with restrictions on the pseudorapidity and transverse momentum
of the $e^\pm$, events with large $|\delta
x_{max(min)}|=|x_{max(min)}^{sim}-x_{max(min)}^{recon}|$ 
still arise.  The reasons for this are:  (1) events with 
sizeable $q_{T}$ are predicted to occur by PYTHIA even when the
daughter $e^\pm$ is detected at $\eta>$0,
(2) the mass distribution of the $W$ can result in events with $M_W$
significantly different from its central value of 80.4 GeV/c$^2$, and (3)
some of the daughter $e^\pm$ can be produced by the decay chain
$W^{\pm}\rightarrow\tau^{\pm}+\nu_{\tau}\rightarrow
e^\pm+\nu_e+\nu_{\tau}$ meaning that the $W$ rest energy is shared
between three, rather than two, final state particles.  Some additional
ambiguity will arise in the assignment of $x_{max(min)}$ to the
interacting $q$ and $\overline{q}$.  However, for most of the events
$x_{max}$ will 
correspond to $x_q$ and $x_{min}$ to $x_{\overline{q}}$.  
To quantitatively assess the efficacy of this kinematics
reconstruction procedure, simulations of $A_L$ using specific models
for quark and antiquark structure functions must be performed.  The
kinematics reconstruction can be used to attempt to directly extract
the quark polarization from the simulated $A_L$ in the appropriate
part of phase space.

\section{Summary}

An exciting program of studies in `spin physics' will commence at RHIC
in 2001, when the first $\vec{p}+\vec{p}$ collisions at high energy
will be observed by the PHENIX and STAR detectors.  
The development of the polarization and intensity
of the colliding proton beams will parallel the construction of the
STAR barrel and endcap electromagnetic calorimeters, and the
implementation of other necessary changes to STAR required for the
measurement of polarization observables in high-luminosity
$\vec{p}+\vec{p}$ collisions.  The present plan is that the RHIC spin
program will be more fully developed by 2003.  Measurements in the
ensuing years will provide important tests of QCD and data that should
help to unravel the spin structure of the proton.  Possibly the most
significant result that will come from the RHIC-spin program is the
determination of the fraction of the proton's spin carried by gluons.




\end{document}